# STATIONARY SOLUTIONS OF THE ELECTROMAGNETIC TE WAVES PROPAGATING ALONG A SINGLE INTERFACE BETWEEN THE TWO KERR-TYPE NONLINEAR MEDIA

M. WABIA  AND  J. ZALEŚNY [*]

Institute of Physics, Technical University of Szczecin
Al. Piastów 48, 70-310 Szczecin, Poland

Abstract:

Propagation of the TE electromagnetic waves in self-focusing medium is governed by the nonlinear Schrödinger equation. In this paper the stationary solutions of this equation have been systematically presented. The phase-plane method, qualitative analysis, and mechanical interpretation of the differential equations are widely used. It is well known that TE waves can be guided by the single interface between two semi-infinite media, providing that one of the media has a self-focusing (Kerr type) nonlinearity. This special solution is called a spatial soliton. In this paper our interests are not restricted to the soliton solutions. In the context of the nonlinear substrate and cladding we have found solutions which could be useful to describe also the incident light in nonlinear medium. This result is the main point of the paper. Some of the presented stationary solutions were already used in similar optical context in literature but we show a little wider class of solutions. In the last section we review and illustrate some results concerning the spatial soliton solution.

## 1. Introduction

Propagation of light in Kerr medium has been already widely investigated in the literature. In this paper we focus our attention on the TE wave propagating along single interface between two Kerr media. Very similar problems have already been examined by many authors [1-9]. It was found that TE waves can be guided by the interface between two semi-infinite media, providing that one of the media has a self-focusing nonlinearity. This is a special solution called the spatial soliton.

In some sense our paper is a review and a extension of those papers. In our paper we assume that both substrate and cladding are two different anisotropic Kerr type nonlinear media. Moreover, we do not want to restrict ourselves to soliton solutions, but to show a whole class of possible stationary solutions existing for this case. Some of the presented stationary solutions were already used in similar context in literature [1-9] but we will show a little wider class of solutions.

In Fig.1 the coordinate system that will be used in this paper is defined. The interface is in the *yz*-plane and propagation is along the *z*-axis with no dependence upon y.

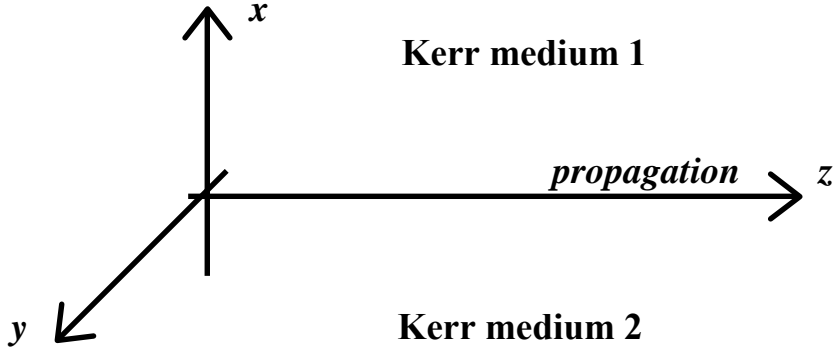

Fig. 1: Geometry of the dielectric structure.

Under this condition (i.e. $\partial/\partial y = 0$) the time dependent Maxwell equations fall into two groups: TE and TM waves, see e.g. [7]. Here we are interested in TE waves only. We also assume harmonic time dependence of fields with a single angular frequency $\omega$. Then TE waves are described by the set of differential equations

$$i\omega \mu_0 \; \underline{H}_z = \partial_x \underline{E}_y,$$

$$i\omega \mu_0 \; \underline{H}_x = -\partial_z \underline{E}_y,$$

$$i\omega \varepsilon \; \underline{E}_y = \partial_x \underline{H}_z - \partial_z \underline{H}_x.$$

(1)

We look for solutions in the form

$$\underline{E}_y(x,y,z) = E_y(x)\,\exp[i(k_z z - \omega t)], \tag{2}$$

$$\underline{H}_x(x,y,z) = H_x(x)\,\exp[i(k_z z - \omega t)],$$

$$\underline{H}_z(x,y,z) = H_z(x)\,\exp[i(k_z z - \omega t)].$$

Our task is to find the functions $H_x(x), H_z(x), E_y(x)$. Because we do not admit any dependence on coordinate $z$, it means that we look for stationary solutions. For simplicity we will omit the coordinate $x$ further on. Considering propagation of light along the optical axis of each medium we reduced the susceptibility tensor to a diagonal form as

$$\varepsilon = \begin{bmatrix} \varepsilon_{11} & & \\ & \varepsilon_{22} & \\ & & \varepsilon_{33} \end{bmatrix} \tag{3}$$

Since we are interested in medium of Kerr type nonlinearity we put for $\varepsilon_{22}$

$$\varepsilon_{22} = \varepsilon + a |E_y|^2, \tag{4}$$

where $\varepsilon$ and $a$ are linear and nonlinear coefficients, respectively. One can easily check that combining equations (1), (2), (3) and (4) the nonlinear equation for field component $E_y(x)$ is obtained

$$d^2 E_y / dx^2 - \kappa^2 E_y + 2\lambda |E_y|^2 E_y = 0, \tag{5}$$

where we use notation

$$\kappa^2 = k_z^2 - k_0^2 \varepsilon, \quad \lambda = \tfrac{1}{2} k_0^2 a, \quad k_0 = \omega/c. \tag{6}$$

For self-focusing medium the following inequalities are fulfilled

$$a > 0, \quad k_z^2 - k_0^2 \varepsilon > 0. \tag{7}$$

Equation (5) is a stationary version of the famous Nonlinear Schrödinger Equation.
In the next sections we briefly discuss its solutions. These solution are very well known for mathematicians but it is interesting to discuss them in this optical context. Our goal is to present a mechanical interpretation of the differential equations and we extentively do it here qualitatively [10].

## 2. Duffing equation

In general the electric field component $E_y$ should be regarded as complex. However, in this section we examine only real $E_y$. Under such an assumption the equation (5) is just the Duffing equation [10] which is known in nonlinear mechanics as one of the simplest nonlinear generalisation of the harmonic oscillator. It is also one of the form of self-interacting scalar field model $\varphi^4$ of field theory

$$d^2 E_y / dx^2 - \kappa^2 E_y + 2\lambda E_y^3 = 0. \tag{8}$$

Equation (8) can be integrated at once to give the first integral

$$(dE_y / dx)^2 - \kappa^2 E_y^2 + \lambda E_y^4 = C. \tag{9}$$

In mechanics the interpretation of the constant of integration $C$ is obvious. It is simply the total energy of the nonlinear oscillator. The first term is the kinetic energy and the next two terms

$$U = -\kappa^2 E_y^2 + \lambda E_y^4 \tag{10}$$

give the potential energy $U$. Contrary to a parabolic potential of harmonic oscillator the potential of Duffing oscillator has an additional term. In Fig.2 we present the plot of $U$.

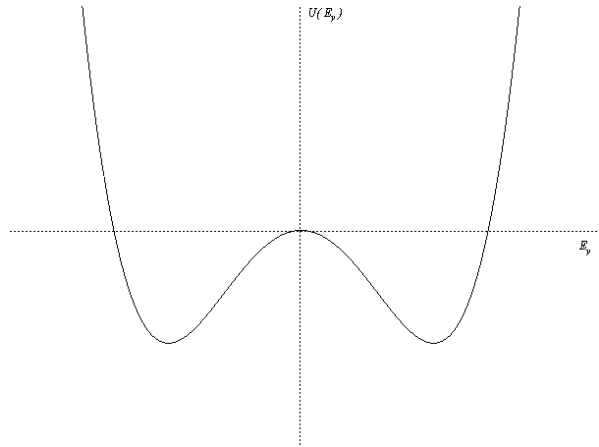

Fig. 2: The 'potential energy' of Duffing oscillator (eq. (10)).

Apart from the potential energy it is useful to plot the 'phase diagram', i.e., trajectories of the Duffing equation (5) in coordinates $(E_y, dE_y/dx)$, see Fig.3. It enables us a qualitative discussion of solutions.

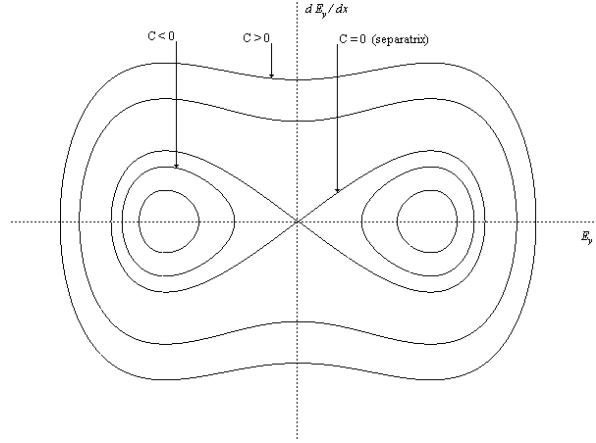

Fig. 3:    Phase diagram of Duffing equation (8).

One can distinguish three different types of solutions.
1. $C < 0$.
*'Small amplitude'* solutions close to the stable points $\pm E_0$. They are inside separatrix curve. Analytically, they are described in terms of elliptic functions. We present the solutions in terms of Jacobi elliptic functions though Weierstrass elliptic functions can also be used here

$$E_y \;=\; [\kappa^2/\lambda\,(2-m^2)]^{1/2}\; dn\{\,[\kappa^2/(2-m^2)]^{1/2}\,(x-x_0\,),\,m\,\} \qquad (11)$$

with the following relation between parameters $m$ and $C$

$$C \;=\; -(\kappa^4/4\lambda)\,(1-m^2)/(1-m^2/2)^2; \qquad 0 \le m < 1. \qquad (12)$$

2. $C = 0$.
*Soliton solution* i.e. exactly separatrix curve. Here Jacobi functions degenerate to hyperbolic ones

$$E_y \;=\; \pm\,(\kappa^2/\lambda)^{1/2}\{\,1/ch[\kappa(x-x_0\,)]\,\}; \qquad m = 1. \qquad (13)$$

3. $C > 0$.
*'Large amplitude'* solutions given again in terms of Jacobi functions

$$E_y \;=\; [\kappa^2/\lambda(2m^2-1)]^{1/2}\,m\; cn\{\,[\kappa^2/(2m^2-1)]^{1/2}\,(x-x_0\,),\,m\,\}, \qquad (14)$$

where the parameters $m$ and $C$ subject the condition

$$C \;=\; (\kappa^4 m^2/\lambda)\,(1-m^2)/(2m^2-1)^2\;; \qquad (1/2)^{1/2} \le m < 1. \qquad (15)$$

Certainly in our optical problem, the interpretation of the constant C is quite different than in mechanics. In particular it is not energy of the system. As we will see later, it is a constant which is connected with boundary conditions.

## 3. Complex case

We have already mention that in general $E_y$ should be considered as complex. In this section we examine this case and find the solutions. To our knowledge these solutions have not been presented in this context yet. We express $E_y(x)$ as

$$E_y(x) = E(x)\ exp[i\ \varphi(x)]. \tag{16}$$

Both functions $E(x)$ and $\varphi(x)$ are real. In this way we admit the dependence of the phase $\varphi$ on $x$ coordinate. Our first motivation to find an exact form of such a solution was to describe a wave which is incident at some angle onto interface. For instance, one can easily notice that equation (5) admits solutions in a plane wave form with any wave vector lying in $xz$ plane. Of course (16) is a generalisation of such plane waves. Inserting (16) into (5) results in a set of two equations

$$d^2E/dx^2 - E(d\varphi/dx)^2 - \kappa^2 E + 2\lambda E^3 = 0, \tag{17}$$

$$E(d^2\varphi/dx^2) + 2(d\varphi/dx)(dE/dx) = 0.$$

The second of them can be easily integrated to give

$$d\varphi/dx = K/E^2, \tag{18}$$

where $K$ is integration constant. This constant refers to energy flux propagating to or from the interface, see, e.g. [7].
The combination of equation (18) with the first of the equations (17) gives

$$d^2E/dx^2 - \kappa^2 E + 2\lambda E^3 - K^2/E^3 = 0. \tag{19}$$

This equation (19) differs from Duffing equation that it contains the term with the constant $K$. There is no problem to find its first integral

$$(dE/dx)^2 - \kappa^2 E^2 + \lambda E^4 + K^2/E^2 = C_\varphi. \tag{20}$$

Thus (20) contains two integration constants K and $C_\varphi$.
Likewise, as in pure Duffing equation case, it is useful to plot phase diagram for this case, in order to discuss the solutions qualitatively, see Fig.4.

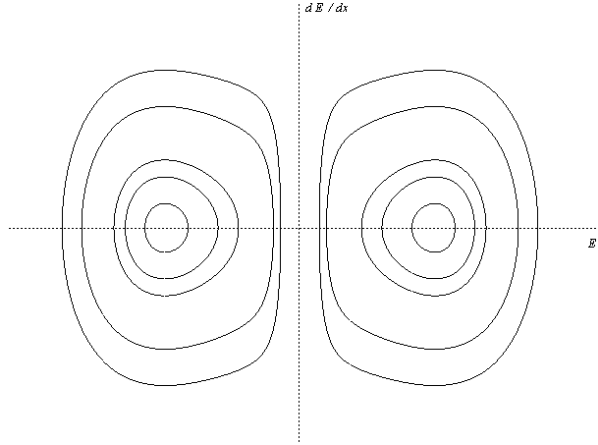

Fig. 4:    Phase diagram of equation (19).

This phase diagram has qualitatively different properties than the one in Fig.3. That is quite naturally explicable in nonlinear oscillator analogy, because the 'potential energy' is (see Fig.5)

$$U = -\kappa^2 E^2 + \lambda E^4 + K^2/E^2 . \tag{21}$$

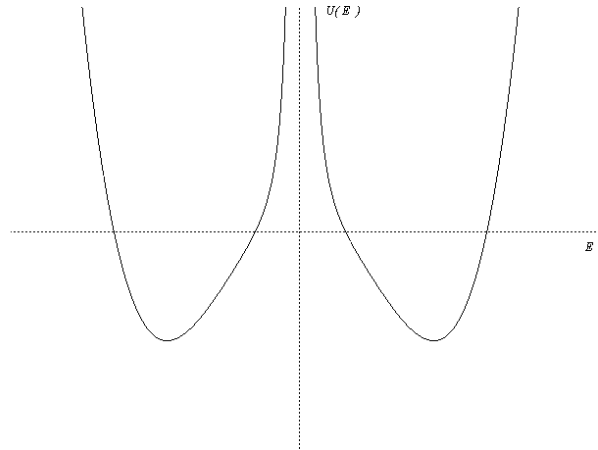

Fig. 5:    The 'potential energy' (21) of the equation (19).

As it is easily seen in Fig.5 ($K \neq 0$ !), there are always two separate potential wells. The first and the most important conclusion which can be drawn from these diagrams is *the lack of soliton solution*. Certainly, it is because of the lack of any separatrix in the phase diagram.
Multiplying the equation (20) by $E^2$, and defining $y \equiv E^2$ we can recast it to the form

$$\tfrac{1}{4}(dy/dx)^2 + \lambda y^3 - \kappa^2 y^2 - C_\varphi y = -K^2 . \tag{22}$$

Once more we can use mechanical analogy and look upon equation (22) as the first integral of a certain nonlinear oscillator (with quadratic nonlinearity in equation of motion). Then the quantity

$$U = \lambda y^3 - \kappa^2 y^2 - C_\varphi y, \qquad (23)$$

plays a role of the 'potential energy' and the constant $-K^2$ simply stands for the 'total energy' of the oscillator. Note that for $K \neq 0$ it is always negative. As before, it is useful to look at the diagram of 'potential energy' (23), see Fig.6.

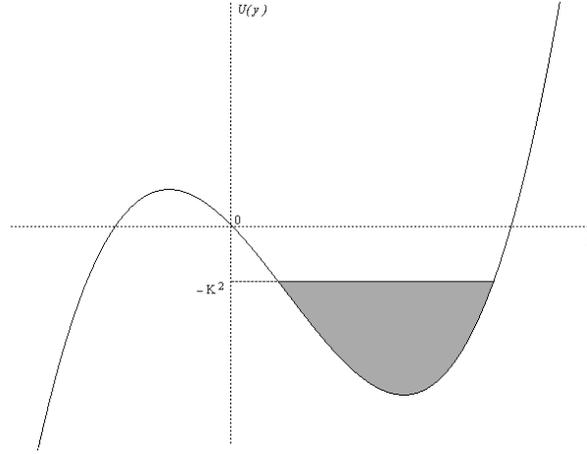

Fig. 6:   The 'potential energy' given by the equation (23).

Keeping in mind that the diagram has the physical sense only for $y > 0$ (because $y \equiv E^2$), one can easily notice that solutions could only be in the form of oscillations inside the potential well (the shaded area in the diagram). It is certainly in full agreement with previous qualitative analysis of the equation (20). Though the equation (22) gives nothing new at this moment, it will be very useful later. A simple conclusion from Fig.6 is the necessary condition for the existence of solutions

$$\lambda y_{min}^3 - \kappa^2 y_{min}^2 - C_\varphi y_{min} \leq -K^2, \qquad (24)$$

where $y_{min}$ is the minimum of the potential well. It does not make any trouble to find the value of $y_{min}$ but since it is not of much interest we skip this here.
The analytical solution of equation (22) is

$$y = \alpha - \beta m^2 sn^2[\mu(x - x_0), m], \qquad (25)$$

where the constants $\alpha, \beta$ are solutions of the algebraic equations

$$\lambda \alpha^3 - \kappa^2 \alpha^2 - C_\varphi \alpha + K^2 = 0, \qquad (26)$$

$$\beta^2 - 3\beta\alpha + 3\alpha^2 + (\kappa^2/\lambda)\beta - 2(\kappa^2/\lambda)\alpha - C_\varphi/\lambda = 0.$$

In addition, because $y > 0$ for every $x$, then $\alpha, \beta$ should fulfil the inequalities

$$\alpha > 0 \quad \text{and} \quad \alpha > \beta m^2 . \tag{27}$$

There is no problem to solve the set (26) algebraically in a standard way.
Knowing $\alpha$, $\beta$ one finds $\mu$ and $m$ to be

$$\begin{aligned} \mu^2 &= \lambda \beta, \\ m^2 &= [(3\lambda\alpha - \kappa^2)/\mu^2] - 1 . \end{aligned} \tag{28}$$

Note that the choice $\alpha = \beta$, implies $K = 0$. Then, the above solution reduces to the *'small amplitude'* case of Duffing equation, i.e. to the equations (11) and (12).
So far we used mechanical analogy to examine equation (22). But this equation has its own meaning in optics. In fact, it determines the energy flux along the direction of propagation, i.e. the $z$ component of Poynting vector $S_z$.

$$S_z = -½ \, Re \, (E_y \, H_x) = ½ \, (c^2 \varepsilon_0 \, k_z/\omega) \, E^2 . \tag{29}$$

Let us now define $1/\gamma \equiv ½ \, (c^2 \varepsilon_0 \, k_z/\omega)$ which allows the equation (22) to be written down as the equation for $S_z$, i.e.,

$$¼ \, \gamma^2 \, (d S_z / d x)^2 + \lambda \gamma^3 \, S_z^3 - \kappa^2 \, \gamma^2 \, S_z^2 - C_\varphi \, \gamma \, S_z + K^2 = 0 . \tag{30}$$

The solution of (30) reads

$$\gamma \, S_z = \alpha - \beta \, m^2 \, sn^2[\mu (x - x_0), m] . \tag{31}$$

Knowing $S_z$, the time-averaged power flow $P$ along $z$ axis in each half space can be obtained. It is the integral over $S_z$ from $x = 0$ (interface) to infinity or minus infinity. In general, such integrals diverge. It should not be considered as a disadvantage for an incident wave, because the conservation of energy flux in the nonlinear medium is guaranteed by the finite value of constant $K$, and we do not need the total power to be finite. On the contrary, in the case of surface or guided waves along interface, we expect the power $P$ to be finite. The only possibility to make these integrals finite is to take

$$\alpha = \beta \quad \text{and} \quad m = 1 . \tag{32}$$

It is of course the soliton solution (13) of previously examined Duffing equation.
Thus, for a single interface between two self-focused nonlinear media, there are no other stationary solutions with finite power flow than spatial solitons.

## 4. Solitons

Solitons are the most important and the most strongly examined stationary solutions of the Duffing equation. Such localized (soliton) solutions in the context of a single interface between two nonlinear media were described for the first time by A.G. Litvak and V.A. Mironov in 1968, [1]. They were extensively discussed in many papers later, e.g. [2-9]. In this section we review the formulas describing the soliton solutions. This simple case serves here as an example of matching conditions on the interface. We leave the more general problem of matching conditions for future considerations. Our choice gives the opportunity to present nice plots of the soliton solutions. This is maybe the most interesting point of this section.
Using equation (13) we can write the solutions in each half-space as

$$E_1 = (\kappa_1^2/\lambda_1)^{1/2} \{ 1/ch[\kappa_1(x - x_1)] \}, \tag{33}$$

for the first medium and

$$E_2 = (\kappa_2^2/\lambda_2)^{1/2} \{ 1/ch[\kappa_2(x - x_2)] \}, \tag{34}$$

for the second medium. Note that the maxima of $E_1$, $E_2$ are for $x_1$, $x_2$ respectively. At $x = 0$, the boundary conditions read

$$E_1(x=0) = E_2(x=0) \quad \text{and} \quad (dE_1/dx)|_{x=0} = (dE_2/dx)|_{x=0}. \tag{35}$$

It proves that the field on the interface (let refer to it as $E_0$), depends only on the constants characterising the media and once they are fixed, the field on the interface remains fixed, independent of the wave number $k_z$ or the power flow along the interface.

$$E_0 = [(\kappa_1^2 - \kappa_2^2)/(\lambda_1 - \lambda_2)]^{1/2} = [2(\varepsilon_2 - \varepsilon_1)/(a_1 - a_2)]^{1/2}. \tag{36}$$

Thus, for real $E_0$, the following conditions must be fulfilled

$$\varepsilon_2 > \varepsilon_1 \quad \text{and} \quad a_1 > a_2 \quad \text{or} \quad \varepsilon_2 < \varepsilon_1 \quad \text{and} \quad a_1 < a_2. \tag{37}$$

They determine the media for which a spatial soliton would exist.
The parameters $x_1$, $x_2$ can be calculated from

$$ch(\kappa_1 x_1) = (\kappa_1^2/\lambda_1)^{1/2}/E_0, \quad ch(\kappa_2 x_2) = (\kappa_2^2/\lambda_2)^{1/2}/E_0. \tag{38}$$

Note that boundary condition for derivatives forces both the values to be the same sign, i.e., both points $x_1$, $x_2$ should lie in the same half-space. Thus, in one medium we have only „a tail" from the first solution and in the second medium „a bigger part" of the other solution. Because the parameters $x_1$, $x_2$ must be both positive or both negative, we should take into consideration each of this cases separately. Fig.7 and Fig.8 show both cases respectively.

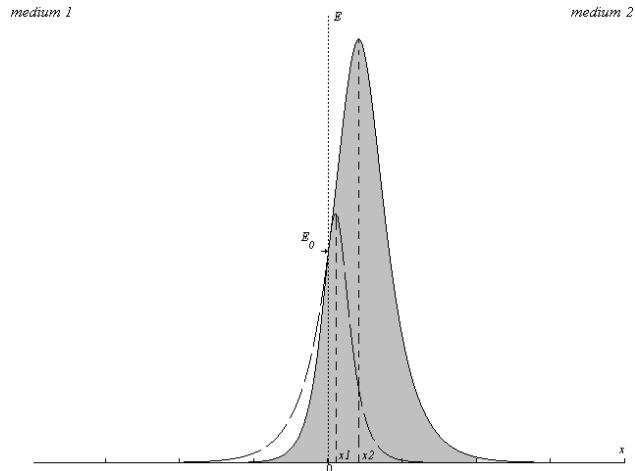

Fig. 7:    The soliton solution (shaded area) in the case of both positive parameters $x_1$, $x_2$.

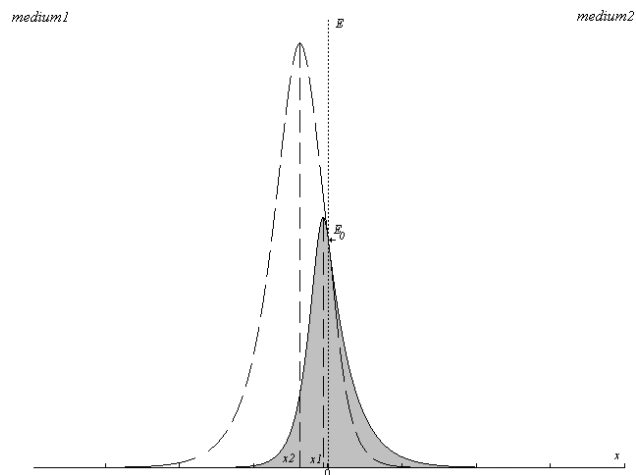

Fig. 8:    The soliton solution (shaded area) in the case of both negative parameters $x_1$, $x_2$.

The proper solutions are marked as shaded area. Henceforth, we will refer to the solution represented by this area simply as a soliton or a soliton solution. Note a significant difference between the two figures. The sizes and shapes are quite different. Summarising, we can have a soliton (the shaded area) in either of the media. In which of them, in fact, depends on boundary conditions, more precisely, it depends on the sign of derivatives in (35).

Apart from the two above cases there is also a remarkable intermediate case, namely, when $x_1 = x_2 = 0$. Then both „soliton peaks" lie on interface in the way shown in Fig.9.

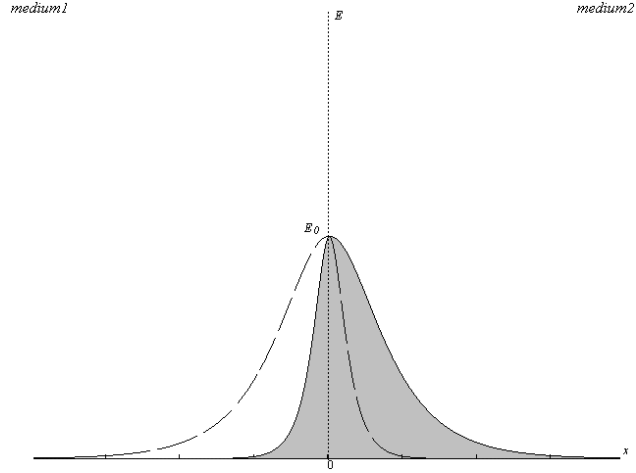

Fig. 9:   The soliton solution (shaded area) in the case when $x_1 = 0$ and $x_2 = 0$.

It is only possible if the propagation constant $k_z$ satisfies the condition

$$\kappa_1^2/\lambda_1 = \kappa_2^2/\lambda_2, \tag{39}$$

or, equivalently

$$k_z^2 = k_0^2 (a_1 \varepsilon_2 - a_2 \varepsilon_1)/(a_1 - a_2). \tag{40}$$

Notice, that amplitude of this soliton is $E_0$. Since it is fixed, then with regard to the amplitude it is the smallest soliton at all. But as we will soon see, it is not the smallest soliton in the meaning of the power flow. The area under the curve $E(x)$ is proportional to the power provided. It could be calculated directly following Boardman *et al*. [ ]. Certainly, we should take into consideration the signs of the parameters $x_1$, $x_2$. Thus, there are two branches of the total power $P_{tot}$. For negative $x_1$, $x_2$ it is

$$P^-_{tot} = (c^2 \varepsilon_0 k_z / \omega k_0^2) \{[(\kappa_1 + (\kappa_1^2 - \lambda_1 E_0^2)^{1/2})/a_1] + [(\kappa_2 - (\kappa_2^2 - \lambda_2 E_0^2)^{1/2})/a_2]\}. \tag{41}$$

And for positive $x_1$, $x_2$ it is

$$P^+_{tot} = (c^2 \varepsilon_0 k_z / \omega k_0^2) \{[(\kappa_1 - (\kappa_1^2 - \lambda_1 E_0^2)^{1/2})/a_1] + [(\kappa_2 + (\kappa_2^2 - \lambda_2 E_0^2)^{1/2})/a_2]\}. \tag{42}$$

Both branches meet each other at propagation constant given by equation (40). We will refer to it as $k_{bi}$. As a matter of fact $k_{bi}$ is a bifurcation point. In Fig.10 both branches of power are plotted versus propagation constant $k_z$. Note that $P^-_{tot}$ has the minimum in $k_{min}$ - the point close but different from $k_{bi}$.

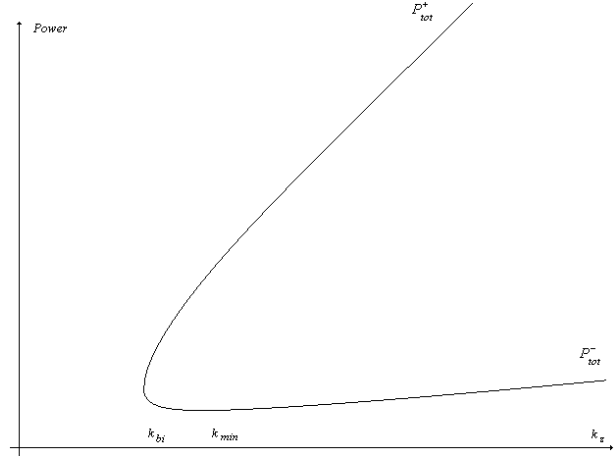

Fig.10: Time-averaged power flow $P_{tot}$ versus propagation constant $k_z$.

To create soliton described by the upper branch $P^+_{tot}$ much more power is required than for soliton described by the branch $P^-_{tot}$. Thus $P^-_{tot}$ should be treated as stable and $P^+_{tot}$ as unstable branch of power. Note also that $k_{bi}$ is a cut-off point and that minimum of power at $k_{min}$ proves the existence of a threshold, i.e., there is a limit of power from below where no soliton occurs.

## 5. Conclusions

In this paper we have systematically presented possible analytic solutions of the stationary version of the Nonlinear Schrödinger Equation (5). We have achieved it using a qualitative analysis of differential equations, i.e., the phase-plane method [10]. We have examined both, the real and the complex cases for the electric field component $E_y$. The main point of our paper are equations (25) and (31). To our knowledge, these solutions have not yet been presented in the context of a single interface between two nonlinear media. We hope that in future investigations such solutions could be useful to describe the incident light in nonlinear medium. As an example of matching conditions on the interface, we have presented, in the last section, a very special case of the solutions, which is a spatial soliton.

# References

[1]   A.G.Litvak, V.A.Mironov, *Izv. VUZ Radiofizika* **11**, 1911 (1968).

[2]   W.J.Tomlinson, *Opt. Lett*. **5**, 323 (1980).

[3]   A.D.Boardman, P.Egan, *IEEE J. Quantum Electron*. **QE-21**, 1701 (1985).

[4]   A.A.Maradudin, „Nonlinear surface electromagnetic waves" in *Optical and Acoustic Waves in Solids-Modern Topics*, World Scientific Pub.,Singapore 1983, Chap. 2.

[5]   A.D.Boardman, P.Egan, „Nonlinear electromagnetic surface and guided waves theory",  in *Proc. 2nd Int. Conf. Surface Waves in Plasmas and Solids*, S. Vukovic, Ed. (Ohrid, Yugoslavia), Singapore, World Scientific Pub. 1986, p.3.

[6]   T.P.Shen, A.A.Maradudin, G.I.Stegeman, „Low-power, single interface guided waves mediated by high-power nonlinear guided waves: TE case", *J. Opt. Soc. Am*. **B 5**, 1391 (1988)

[7]   A.D. Boardman, P.E. Egan, T. Twardowski and M. Wilkins, „Nonlinear surface-guided waves in self-focusing optical media" in *Nonlinear Waves in Solid State Physics*, Vol. 247 of NATO Advanced Study Institute, Series B: Physics, edited by A.D. Boardman et al., Plenum Press, New York 1990, p.1.

[8]   A.D. Boardman, P.E. Egan, F. Lederer, U. Langbein, D. Mihalache „Third-order nonlinear electromagnetic TE and TM guided waves" in *Nonlinear Surface Electromagnetic Phenomena*, edited by H.E. Ponath and G.I. Stegeman, North-Holland, Amsterdam, 1991, p.73.

[9]   E.Wright, G.I.Stegeman, „Nonlinear planar waveguides" in *Anisotropic and Nonlinear Optical Waveguides*, edited by C.G. Someda and G. Stegeman, Elsevier Science Pub. B.V., New York, 1992, p.117.

[10]  A.M.Kosevich, A.S.Kovalev, An Introduction to Nonlinear Physical Mechanics, Kiev, Nauk. dumka, 1989. (Russian)